# A floating body with no preferred orientation: an experimental realization

Lucie Pontiggia[†], Angélique Campaniello[†] and Emmanuel Fort[*].

*Institut Langevin, ESPCI Paris, PSL University, CNRS, 1 rue Jussieu, 75005 Paris, France.*

[*] Corresponding author. Email: emmanuel.fort@espci.fr

[†] These authors contributed equally to this work.

**Abstract**

We present a simple experimental realization of a two-dimensional floating body that can remain in equilibrium in any orientation. This system is based on a class of shapes known as Zindler curves, which possess the remarkable geometric property that all chords dividing their area into equal parts have the same length.

Using a multilayer fabrication approach, we construct a heart-shaped floating object with an effective density close to one half of that of the surrounding liquid. We show experimentally that, under these conditions, the object exhibits neutral equilibrium with respect to rotation. When the density is slightly varied, preferred orientations emerge, consistent with a simple energy-based description.

Our experiments highlight both the accessibility of this classical problem and the subtle role of physical effects such as density inhomogeneities and capillarity. They provide a simple platform to explore the interplay between geometry and buoyancy, and to test geometric results in a tangible setting.

**Introduction**

The flotation of bodies is one of the oldest problems in fluid mechanics. Since Archimedes, it has been known that an object immersed in a fluid experiences an upward force equal to the weight of the displaced fluid. A more subtle question, however, concerns the orientation of a floating object: why do some bodies float in a well-defined orientation, while others appear able to adopt multiple equilibrium positions?

In hydrostatic equilibrium, the buoyant force acts at the center of buoyancy, defined as the center of mass of the immersed volume. Equilibrium then requires that the center of mass of the body and the center of buoyancy be vertically aligned. For most objects of arbitrary shape, this condition can only be satisfied for a limited number of orientations, some stable and others unstable. This naturally raises the following question: do there exist bodies that can float in equilibrium in all orientations?

This question is associated with a well-known mathematical problem posed by Stanisław Ulam in the *Scottish Book* of the Lwów School [1]. In its simplest form, the question asks: is the sphere the only homogeneous solid that can float in equilibrium in any orientation? This problem, now known as the Ulam's floating body problem, has attracted considerable interest in both geometry and the mechanics of floating bodies [2–4].

In two dimensions—corresponding to finding the cross-section of a cylinder floating at the surface of a liquid—the answer has long been known. As early as 1938, Auerbach showed that, for a relative density $\rho = 1/2$, there exist non-circular shapes that can float in all orientations [5]. These shapes are bounded by special curves, now known as Zindler curves [6].

These curves possess a remarkable and, at first glance, surprising geometric property: all chords that divide the area into two equal parts have the same length. This property ensures that the waterline can rotate around the body without altering the condition of hydrostatic equilibrium.

Since these pioneering works, an extensive mathematical literature has developed around neutral floating bodies and Zindler curves, revealing unexpected connections with other geometric and mechanical problems [7–10]. Significant progress has also been made recently in the general N-dimensional case, where the existence of bodies floating in all orientations has been established in some cases, although their explicit shapes remain unknown [11–13].

It is the recent revival of interest in this problem, in the context of these modern developments, that motivated us to revisit it. We were struck both by the simplicity and the counterintuitive nature of the underlying geometric property, and by the fact that it appeared amenable to a simple tabletop experiment. Surprisingly, however, experimental realizations of such objects seem to be absent from the literature.

In this article, we present an experimental realization of a two-dimensional floating body in the spirit of Ulam's problem, based on a simple Zindler curve with the playful shape of a heart. We show that such an object, when its relative density is close to $\rho = 1/2$, can indeed float in equilibrium in all orientations.

We first derive the geometric property that explains this behavior, and then describe the experimental approach and the results obtained. Beyond this qualitative verification, we also investigate what happens when the density of the object slightly deviates from the critical value $\rho = 1/2$. In this case, certain orientations become energetically favorable, while others become unstable. The analysis of these configurations naturally introduces the notion of an effective potential energy landscape associated with the floating of a body of given shape.

Our goal is simply to show how an elegant geometric idea can be turned into a simple and visually striking experiment, capable of illustrating in a pedagogical way the principles of buoyancy and equilibrium in fluids.

**Theory: Geometric condition for orientation-independent floating**

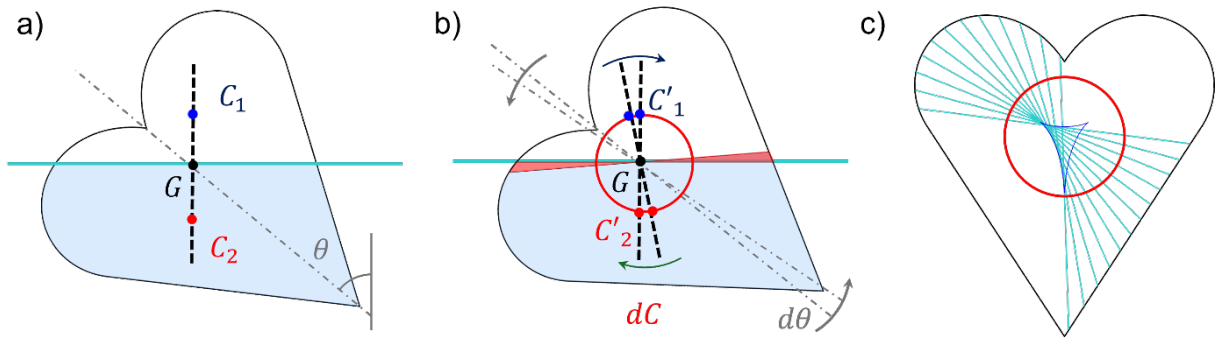

***Figure 1: Geometric construction of orientation-independent floating.*** *(a) Schematic of a two-dimensional floating object with relative density $\alpha = 0.5$. $G$, $C_1$, and $C_2$ denote the centers of mass of the whole object, its emerged part, and its immersed part, respectively. The angle $\theta$ defines the orientation of the object. (b) Infinitesimal rotation $d\theta$ of the object. The immersed region (center of mass $C_2'$) is modified through an exchange of two equal triangular areas along the waterline (red) with the emerged region (center of mass $C_1'$). As $\theta$ varies, $C_1'$ and $C_2'$ move along a circle (red) centered on $G$. (c) Example of a Zindler shape. Several waterlines $L(\theta)$ are shown together with their envelope (blue), and the locus of the center of mass of the immersed region (red circle).*

To analyze the stability of the object as a function of its orientation, we adopt an energy-based description of the problem. The idea is to compare the floating configuration to a simple reference state in which the object is placed next to the liquid, at the same height as the free surface, before being immersed.

Within this framework, immersing the object in the liquid amounts to replacing a volume of fluid by the solid. The potential energy of the combined system object-fluid can then be obtained by summing the gravitational contributions of the solid and subtracting that of the displaced fluid.

Figure 1a introduces the relevant quantities. We consider a homogeneous object of planar area $A$, with center of mass $G$. The waterline divides the object into an emerged region of area $A_1$, with center of mass $C_1$, and an immersed region of area $A_2$, with center of mass $C_2$. We denote by $h_1(\theta)$ and $h_2(\theta)$ the vertical positions of these centers of mass relative to the waterline. The densities of the object and the liquid are $\rho_{\text{obj}}$ and $\rho_{\text{liq}}$, respectively.

The gravitational energy of the solid can be written as the sum of the contributions of the emerged and immersed parts, $E_{\text{solid}} = \rho_{\text{obj}} g A_1 D\, h_1(\theta) + \rho_{\text{obj}} g A_2 D\, h_2(\theta)$, where $D$ is the thickness of the object. This must be compared to the energy of the fluid that would occupy the immersed region in the absence of the object. This displaced fluid, of density $\rho_{\text{liq}}$, would contribute $E_{\text{fluid}} = \rho_{\text{liq}} g A_2 D\, h_2(\theta)$. The potential energy of the floating object yields $E_p(\theta) = \rho_{\text{obj}} g A_1 D\, h_1(\theta) + \rho_{\text{obj}} g A_2 D\, h_2(\theta) - \rho_{\text{liq}} g A_2 D\, h_2(\theta)$.

Introducing the relative density $\alpha = \rho_{\text{obj}}/\rho_{\text{liq}}$ and using the relations $A_1 = (1-\alpha)A$ and $A_2 = \alpha A$, we obtain

$$E_p(\theta) = \rho_{\text{obj}} g (1-\alpha) A D\, [h_1(\theta) - h_2(\theta)]. \qquad (1)$$

This expression is particularly insightful, as all the dependence on the orientation is contained in the height difference between $C_1$ and $C_2$. Since $G$ is the barycenter of $C_1$ and $C_2$ with constant weights, the vertical distance between $C_2$ and $G$ remains constant. As the object rotates, the center of mass $G$ remains fixed within the body. If the object floats in equilibrium for all orientations $\theta$, the potential energy must be independent of $\theta$, which implies that $C_2$ must always lie vertically below $G$. In other words, $C_2$ moves along a circle centered at $G$.

To relate this condition to a geometric property of the shape, we consider an infinitesimal rotation $d\theta$ of the object. The waterline then rotates slightly, modifying the immersed region. To leading order, this variation can be described as an exchange of two small triangular regions, one emerging from the liquid and the other becoming immersed (see Fig. 1b). Since $C_2$ is the barycenter of the immersed region, its displacement $\delta C_2$ induced by the rotation can be determined from the properties of these triangles.

Hydrostatic equilibrium requires that the immersed area remains constant, $A_2 = \alpha A$, independently of the orientation. Therefore, the area variation must vanish, implying that the two triangles have equal area. This transformation is thus equivalent to a rotation about the midpoint of the waterline. To leading order, the area of each triangle scales as $L(\theta)^2 \mathrm{d}\theta/8$, where $L(\theta)$ is the length of the waterline. Their centroids lie along the median at a distance $L(\theta)/6$ from the base. This leads to a displacement $\delta C_2 \sim (L(\theta)^3/8\alpha A)\,\mathrm{d}\theta$. Since $C_2$ moves along a circle, its displacement must be proportional to $\mathrm{d}\theta$. This implies that $L(\theta)^3$, and therefore $L(\theta)$, must be independent of $\theta$.

The particular case $\alpha = 1/2$ exhibits a remarkable symmetry. In this situation, the emerged and immersed areas are equal ($A_1 = A_2$), so exchanging these two regions leaves the configuration unchanged. As a result, an equilibrium configuration at an angle $\theta$ corresponds to another equilibrium configuration at $\theta + \pi$. This symmetry has several important consequences. First, the center of mass of the solid coincides with the midpoint of the segment joining $C_1$ and $C_2$, and therefore lies on the waterline for any equilibrium orientation. Second, this invariance significantly reduces the geometric constraints of the problem, as conditions at angles $\theta$ and $\theta + \pi$ are no longer independent.

This reduction allows for a rich family of nontrivial solutions. In particular, the problem can be reformulated in purely geometric terms by considering all chords that divide the area into two equal parts. Curves for which all such chords have the same length form a class of solutions known as Zindler curves [4].

To illustrate these properties, we consider in the following a particular member of this family: a heart-shaped curve introduced by Auerbach [3]. Figure 1c shows several waterlines corresponding to different orientations. Each line divides the shape into two equal areas and therefore corresponds to an equilibrium configuration. The midpoints of these waterlines trace a caustic curve, which forms the envelope of the family of waterlines.

**Experimental realization**

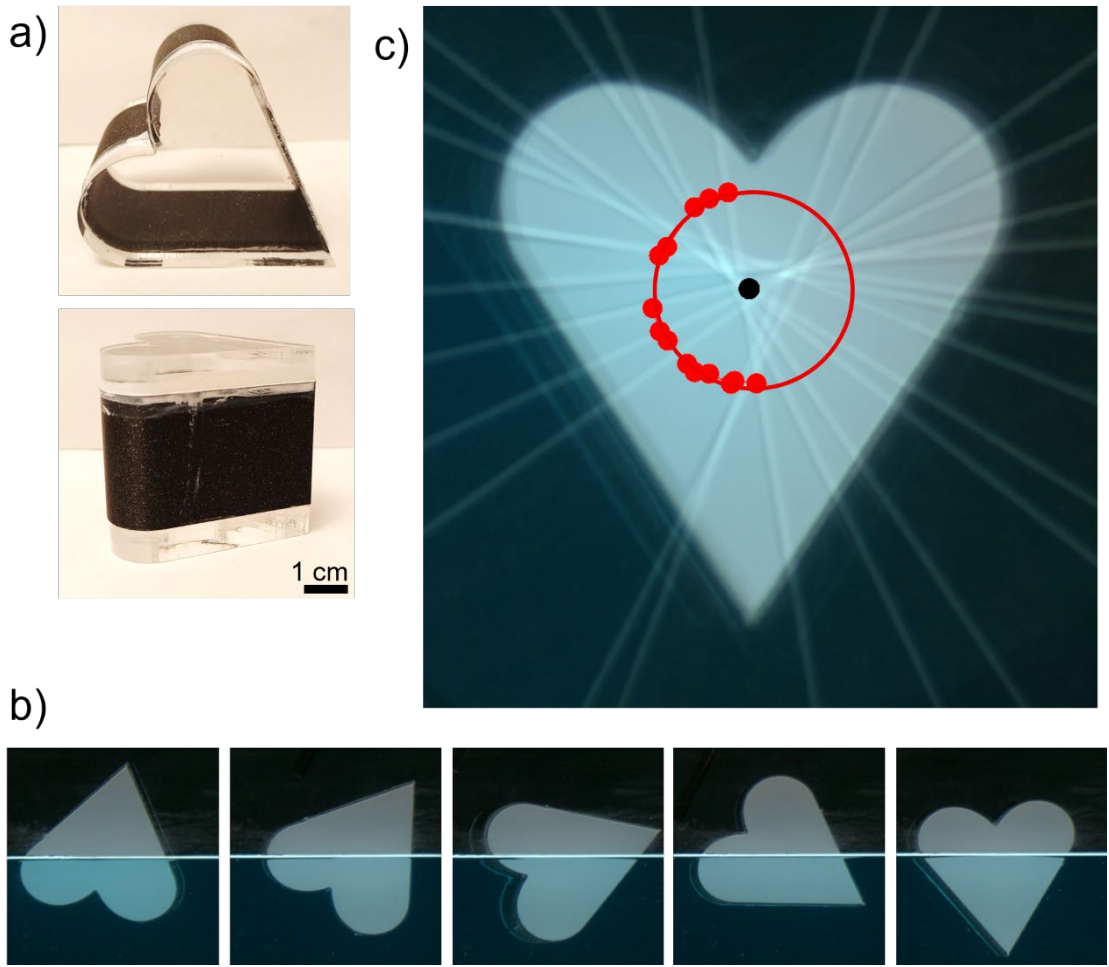

*Figure 2: Experimental realization of a Zindler-shaped floating body. (a) Photograph of the heart-shaped object fabricated using a thin black 3D-printed profile sandwiched between two transparent PMMA plates. This construction allows control of the effective density while maintaining good homogeneity. (b) Side-view images of the floating object in different orientations (see Movie 2). The object remains in equilibrium for arbitrary orientations, illustrating neutral rotational stability. (c) Superposition of the configurations shown in (b). The waterlines appear as bright segments, while the corresponding centers of mass of the immersed regions are shown as red dots. These points lie on a circle centered on the center of mass of the object (black dot). The envelope of the waterlines forms a triangular caustic structure, in agreement with the geometrical construction shown in Fig. 1c.*

We first attempted to fabricate homogeneous objects directly using a 3D printer, by reproducing the heart shape as a cylinder of constant cross-section. The goal was to tune the effective density of the material by adjusting the printing parameters (infill ratio, internal structure) in order to reach a relative density $\alpha = 0.5$. However, despite numerous attempts and different filling strategies, the resulting objects proved insufficiently homogeneous. Even for simple shapes such as circular cylinders, it was not possible to achieve neutral equilibrium in all orientations, highlighting the strong sensitivity of the problem to small density inhomogeneities.

We therefore adopted an alternative approach consisting of fabricating only the outer contour of the shape as a thin structure (typically <0.5 mm thick), and sandwiching it between two transparent PMMA plates. An example is shown in Fig. 2a. The central part, produced by 3D printing, appears dark, while the two transparent plates are bonded on either side. This geometry provides an effective realization of a two-dimensional homogeneous object.

The effective density of the floater can be readily controlled by adjusting the relative thicknesses of the different layers. The PMMA plates (typical density $\rho \approx 1.18 \text{ g.cm}^{-3}$) enclose a printed structure with a much lower effective density, allowing fine tuning of the average density of the assembly. The resulting objects have typical dimensions of a few centimeters and masses of a few tens of grams. Their density is measured experimentally from their mass and volume, yielding values around $\rho_{\text{obj}} \approx 0.5$ with a relative uncertainty of about 0.01.

In practice, the condition on $\alpha$ is particularly critical: small deviations from $\alpha = 0.5$ significantly affect the behavior of the floater and lead to the appearance of preferred orientations, as discussed below. To achieve precise control of the relative density, we adjust the density of the liquid after fabrication of the objects. The liquid consists of a mixture of water ($\rho \approx 1.0 \text{ g.cm}^{-3}$ and ethanol ($\rho \approx 0.79 \text{ g.cm}^{-3}$), whose proportions are tuned to match the desired value of $\alpha$. In practice, the floater density is chosen slightly below 0.5, allowing fine adjustment of the liquid density to reach $\alpha = 0.5$. This method provides a simple and reproducible way to approach the critical condition.

The experiments are performed in a tank with a dark background to enhance contrast. The objects are imaged using a camera placed laterally at a large distance, in order to minimize parallax effects. Illumination is provided by LED panels placed on the sides, while avoiding spurious reflections from the tank, which is a glass cube of side length 10 cm.

The different orientations of the floater are recorded, enabling both qualitative and quantitative analysis of the equilibrium conditions as a function of the relative density. An image-processing routine is used to extract the contour of the object and detect the waterline (see Supplementary Movie 1). From this, we determine the positions of the center of mass $G$, the centers of mass of the immersed and emerged regions ($C_2$ and $C_1$), the inclination angle $\theta$, and the length of the waterline $L$.

**Orientation-independent floating**

Figure 2b shows the floating object in different orientations for a theoretical relative density $\alpha = 1/2$. In practice, the experimental density required to observe this behavior is slightly lower, typically $\alpha \approx 0.49$ (this discrepancy is discussed in the Discussion section). The different configurations are obtained by gently perturbing the orientation of the object using thin rods and then releasing it, as shown in Movie 2.

Remarkably, and counterintuitively, the object does not return to a preferred orientation: it remains in the imposed position, illustrating the absence of a stable equilibrium direction. This behavior is particularly striking during manipulation, as one would naturally expect the object to relax toward a unique equilibrium orientation.

From these images, we developed an analysis routine to extract the contour of the object and determine, for each orientation, the position of the waterline and the center of buoyancy. Figure 2c is obtained by superimposing these configurations. The waterlines appear as bright segments, while the reconstructed centers of buoyancy trace a circular trajectory (in practice a semicircle, by symmetry). A triangular structure also emerges, corresponding to the envelope of the waterlines, i.e., the caustic associated with this family of lines.

A quantitative analysis of the data confirms these observations. The length of the waterline remains nearly constant, with a mean value of 40.4 mm and variations smaller than $\pm 0.9$ mm close to the theoretical value of 4.00 cm. Similarly, the center of mass of the object coincides with the waterline to within 1 mm, in agreement with the theoretical prediction for $\alpha = 1/2$.

**Deviation from the critical density**

To explore the effects of deviations from the critical condition $\alpha = 0.5$, we performed experiments for slightly different relative densities, typically $\alpha = 0.45$ and $\alpha = 0.55$. The potential energy profile is computed from Eq. (1) as a function of the orientation $\theta$. Vertical hydrostatic equilibrium is assumed to be satisfied, which imposes that the emerged and immersed areas adjust so that the buoyant force balances the weight of the object. Under this condition, the potential energy $E_p(\theta)$ depends only on the orientation. It is evaluated numerically by determining, for each angle $\theta$, the positions of the centers of mass of the emerged and immersed regions from the geometry of the heart shape.

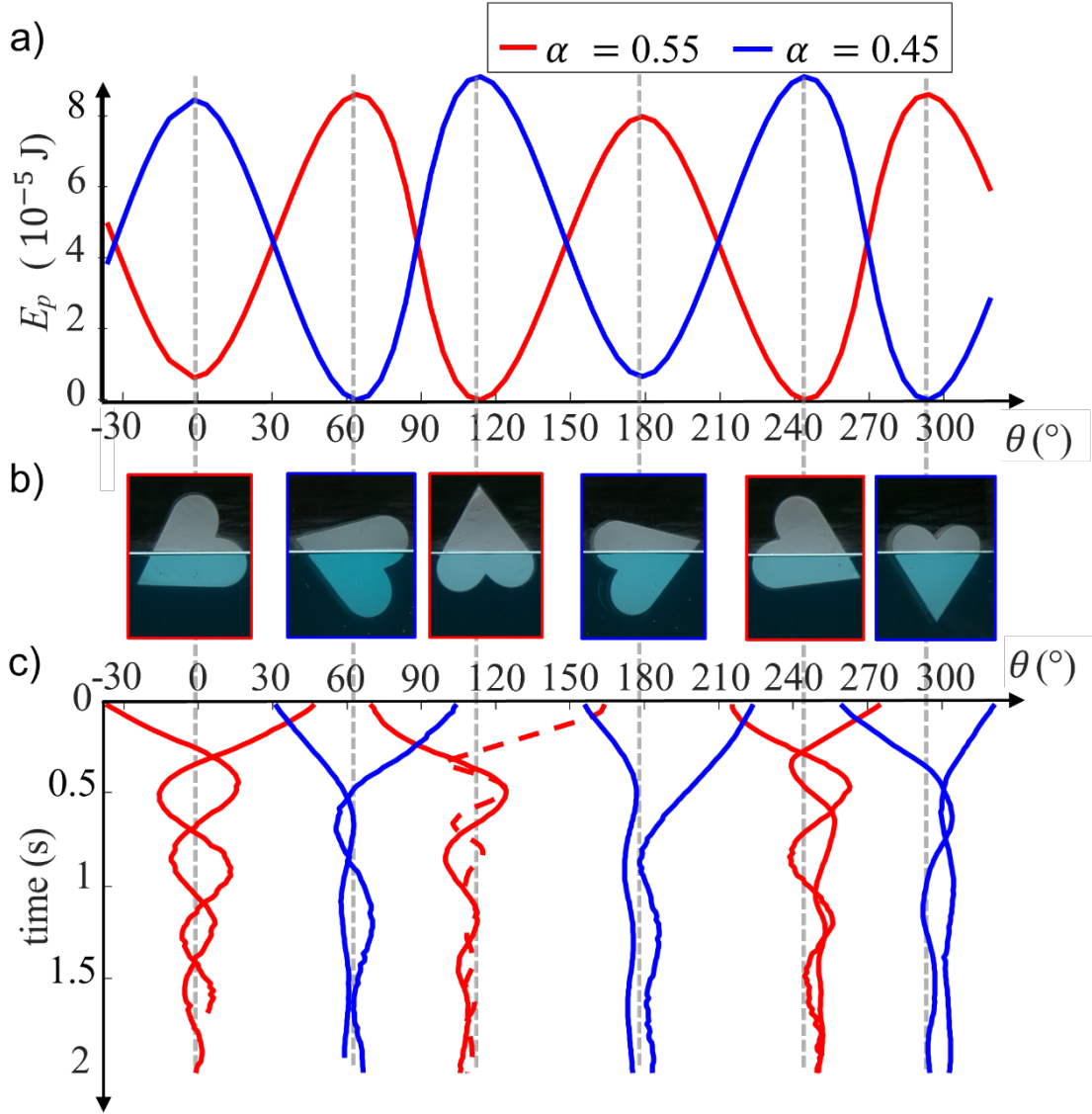

*Figure 3: Energy landscape and dynamics away from the critical density. (a) Potential energy $E_p(\theta)$ as a function of orientation for a denser object ($\alpha = 0.55$, red curve) and a lighter object ($\alpha = 0.45$, blue curve). The minimum energy is set to zero. (b) Experimental images of the floating objects at the corresponding equilibrium orientations. (c) Time evolution of the orientation $\theta(t)$ after release from a nonequilibrium configuration (see Supplementary Movies 3 and 4). The red dashed curve corresponds to a trajectory with a higher oscillation frequency (≈ 4 Hz).*

The energy profiles are shown in Fig. 3a. For $\alpha = 1/2$, the energy is constant (black curve), reflecting the absence of a preferred orientation. In contrast, for $\alpha = 0.45$ and $\alpha = 0.55$, a clear modulation with orientation appears, revealing the existence of three stable equilibrium positions separated by approximately 60°. The two curves are out of phase: orientations that are stable for one density become unstable for the other. Figure 3b shows the floating objects in the corresponding equilibrium orientations for both the denser and lighter cases.

This energy landscape directly governs the dynamics of the floater when released from an out-of-equilibrium configuration. Figure 3c shows the time evolution of the orientation when the object is released near an equilibrium position (see Movie 3). In practice, it is difficult to impose a purely rotational perturbation without slightly disturbing the vertical equilibrium. Nevertheless, the observed trajectories clearly show that the floater relaxes toward the minima of the potential energy through damped oscillations. The final orientations correspond to the stable equilibria predicted by the energy profiles.

For the less dense floaters, small deviations of the final orientation, on the order of 10°, can be observed depending on the initial release conditions. By fitting the trajectories with a damped harmonic oscillator model, we extract characteristic frequencies of $1.15 \pm 0.15$ Hz and $1.50 \pm 0.35$ Hz for the lighter and denser cases respectively. An exception is observed for the denser floater released near 165° (see red dashed curve), which relaxes toward $\sim 110°$ with a higher frequency of approximately 4 Hz (see Fig. 3c and Movie 4). The damping rates are approximately 1 s$^{-1}$.

**Discussion**

The damped oscillations observed around the stable equilibrium positions provide a dynamical validation of the interpretation in terms of a potential energy landscape. Expanding the potential energy $E_p(\theta)$ around a minimum $\theta_0$, one obtains a quadratic approximation of the form $E_p(\theta) \simeq E_p(\theta_0) + \frac{1}{2} k_\theta (\theta - \theta_0)^2$, where $k_\theta = \left.\frac{d^2 E_p}{d\theta^2}\right|_{\theta_0}$ plays the role of an angular stiffness. The dynamics of the floater is then that of a damped harmonic oscillator, $I\ddot{\theta} + c\dot{\theta} + k_\theta(\theta - \theta_0) = 0$, where $I$ is the moment of inertia of the object about its center of mass and $c$ an effective damping coefficient. The corresponding natural frequency is given by $\omega_0 = \sqrt{k_\theta/I}$.

The moment of inertia can be estimated directly from the geometry and mass of the object, allowing a quantitative comparison between the measured frequency and the local curvature of the potential. The estimated oscillation frequencies are in the range between 1 and 1.5 Hz which is in good agreement with the experimentally extracted frequencies.

The damping observed in the experimental trajectories can be attributed to viscous dissipation in the surrounding fluid. The rotation of the floater induces shear and fluid motion, leading to energy loss and exponential decay of the oscillations. Additional contributions may arise from effects at the contact line and from the meniscus. As mentioned previously, it is experimentally difficult to excite a purely rotational mode upon release without inducing a slight vertical perturbation of the floater. The dependence of the potential energy on the vertical position can be computed numerically from Eq. (1). Evaluating the shape of the potential well as a function of vertical displacement $z$ reveals a stiffer dynamic, associated with a higher oscillation frequency, typically of the order of 4 Hz. This frequency likely corresponds to the one observed experimentally for the dashed trajectory in Fig. 3c, as suggested by Supplementary Movie S3.

Surface tension effects associated with the meniscus constitute a natural source of deviation between the ideal theoretical description and the experimental observations. The waterline is indeed the location of a capillary force whose vertical component tends to pull the floater downward. The magnitude of this force is of order $\gamma \sin \phi$, where $\phi$ is the angle of the meniscus with respect to the horizontal, integrated along the contact line. The resulting total vertical force can be estimated as $F_\gamma \sim 2\gamma(L + D)\sin\phi$. For typical values ($\gamma \sim 5 \times 10^{-2}$ N/m, $2(L + D) \sim$ 0.1 m, $\sin \phi \sim 0.5$), this force amounts to about 1% of the weight of the floater. This likely explains why the experimental condition for neutral equilibrium is reached for a slightly lower relative density, typically $\alpha \simeq 0.49$.

In principle, this capillary contribution depends on the orientation of the floater through the local geometry of the contact line and variations of the meniscus angle. However, in the particular case $\alpha = 1/2$, the length of the waterline is independent of orientation. As a result, only variations associated with the contact angles across the thickness of the object remain, which helps preserve the orientation-independent floating behavior despite the presence of surface tension effects.

Surface tension can also generate a torque on the floater if the vertical components of the capillary forces differ on the two sides of the contact line. Denoting by $\phi_L$ and $\phi_R$ the meniscus angles on each side of the object of thickness $D$, the corresponding capillary torque can be estimated as $\tau_\gamma \sim \frac{L}{2} \gamma D \Delta(\sin \phi)$, where $\Delta(\sin \phi) = \sin \phi_R - \sin \phi_L$. For typical values ($\gamma \sim 5 \times 10^{-2}$ N/m, $L \sim D \sim 4 \times 10^{-2}$ m, $\Delta(\sin \phi) \sim 0.5$), one obtains a torque of order $10^{-5}$ N m. This value remains small compared to the torques associated with the potential wells for $\alpha = 0.45$ and $\alpha = 0.55$, which explains why capillary effects do not significantly alter the dynamics in these cases.

However, near $\alpha = 0.5$, where the energy landscape becomes nearly flat, even a small residual capillary torque could in principle favor certain orientations. The absence of a clear spontaneous reorientation suggests that another mechanism is at play, most likely contact line pinning. In this situation, the meniscus does not evolve reversibly, and the contact line remains pinned as long as the applied torque stays below a depinning threshold. As a result, even in the presence of a small residual capillary torque, the floater can remain in an arbitrary orientation.

This interpretation is consistent with the dispersion of the final angles observed for the lighter floaters. Depending on the release conditions, the trajectories converge to slightly different final orientations within a range of about 10°. This corresponds to a typical angular deviation from equilibrium of $\delta\theta \sim 5°$. Using the quadratic expansion of the energy potential, the angular stiffness can be estimated as $k_\theta \sim 2\delta E/(\delta\theta)^2$, leading to a characteristic torque $\tau_{\text{pin}} \sim k_\theta\, \delta\theta$. Using the energy variations based on eq. (1) (see Fig. 3a), this yields $\tau_{\text{pin}}$ of the order of $10^{-5}$ N m, which is comparable to the estimated capillary torque.

This explains why, even in the presence of a small residual capillary torque, the floater can remain in an arbitrary orientation: contact line pinning sets a threshold below which no motion occurs.

**Conclusion**

In this work, we set out to experimentally realize a simple and elegant geometric idea arising from Ulam's floating body problem. By fabricating a floater inspired by a Zindler curve, we have shown that it is possible to obtain an object that can float in equilibrium in all orientations when its relative density is close to $1/2$.

This experiment highlights the inevitable gap between the ideal model and reality. Achieving a truly homogeneous two-dimensional object proves challenging, and additional physical effects naturally arise, in particular capillary forces and the associated torques. Nevertheless, these effects remain sufficiently small that they do not destroy the essential property of orientation-independent floating, underscoring the robustness of the underlying geometric condition.

The experimental approach developed here, based on a sandwich structure that allows fine control of both the density of the floater and that of the liquid, provides a simple and flexible platform for exploring such systems. In particular, it enables access to a wide and well-controlled range of effective densities, which would otherwise be difficult to achieve.

Beyond the specific case studied here, many perspectives remain open. It would be of interest to investigate other Zindler curves, as well as configurations corresponding to densities different from $1/2$, for which additional and potentially rich behaviors have been predicted.

**Acknowledgements**

LP acknowledges internship funding from PSL University UROP program. AC is supported by the project Localization of Waves of the Simons Foundation (Grant No. 1027116). EF is supported by the AXA research fund. The authors acknowledge the use of the PClab facility of ESPCI Paris.

**Conflict of interest**

The authors have no conflicts to disclose.